\documentclass[a4paper,12pt]{article}

\usepackage{amsmath, amsthm, amsfonts, amssymb, color}

\theoremstyle{plain}
\newtheorem{theorem}{Theorem}[section]

\newtheorem{lemma}{Lemma}[section]

\newtheorem{corollary}[theorem]{Corollary}

\theoremstyle{remark}

\numberwithin{equation}{section}

\def\Om{\Omega}

\def\e{\varepsilon}
\def\g{\gamma}

\def\l{\lambda}
\def\p{\partial}
\def\D{\Delta}

\def\a{\alpha}
\def\b{\beta}

\def\si{\sigma}

\def\d{\delta}
\def\L{\Lambda}
\def\z{\zeta}

\def\vp{\varphi}

\def\Odr{\mathcal{O}}
\def\H{W_2}
\def\Hloc{W_{2,loc}}

\def\Hinf{W_\infty}

\def\di{\,\mathrm{d}}

\def\I{\mathrm{I}}
\def\iu{\mathrm{i}}

\DeclareMathOperator{\RE}{Re}
\DeclareMathOperator{\IM}{Im}
\DeclareMathOperator{\spec}{\sigma}

\DeclareMathOperator{\supp}{supp}

\DeclareMathOperator{\rank}{rank}

\DeclareMathOperator{\rot}{rot}

\DeclareMathOperator*{\esssup}{ess\,sup}

\begin{document}

\title{\textbf{Asymptotics for the solutions of elliptic
systems with fast oscillating coefficients}}

\author{D.~Borisov}
\date{}
\maketitle

\vspace{-1 true cm}

\begin{quote}
{\small {\em Nuclear Physics Institute, Academy of Sciences,
25068 \v Re\v z
\\
near Prague, Czechia
\\
Bashkir State Pedagogical University, October Revolution St.~3a,
\\
450000 Ufa, Russia
\\
E-mail: \texttt{borisovdi@yandex.ru}}}
\end{quote}

\begin{abstract}
We consider a singularly perturbed second order elliptic system
in the whole space. The coefficients of the systems fast
oscillate and depend both of slow and fast variables. We obtain
the homogenized operator and in the uniform norm sense we
construct the leading terms of the asymptotics expansion for the
resolvent of the operator described by the system. The
convergence of the spectrum is established. The convergence of
the spectrum is established. The examples are given.
\end{abstract}

\section*{Introduction}

There are many works devoted to the homogenization of the
differential operators in bounded domains with fast oscillating
coefficients (see, for instance \cite{PChSh}--\cite{OIS}).
Similar questions for the operators in unbounded domains are
studied essentially less. At the same time, during last years
the case of an unbounded domain is studied intensively. In the
series of papers \cite{Bi}--\cite{BS5} M.Sh.~Birman and
T.A.~Suslina developed a new original technique which allowed
them to prove the convergence theorem, to obtain the precise in
order estimates for the rates of convergence, and to construct
the first terms in the expansion for the resolvent of a wide
class of differential operators in unbounded domains with fast
oscillating coefficients. It should be stressed that these
results were obtained in the uniform norm sense, while usually
the results for the bounded domains were formulated in the sense
of strong or weak convergence. The approach of M.Sh.~Birman and
T.A.~Suslina is based on the spectral theory and treats the
homogenization as a threshold phenomenon. It is applicable to
the operators those can be factorized, and at the same time
their coefficients must depend of the fast variable $x/\e$ only;
the dependence of the slow variable $x$ is not allowed. We
should also note the paper of V.V.~Zhikov \cite{Zh3}, where by
employing another technique he obtained the precise in order
estimates for the rate of convergence for the resolvent of a
scalar operator as well as for the operator of the elasticity
theory. It was assumed that the coefficients are periodic and
depend of the fast variable only, too.

The one-dimensional scalar operators with the coefficients
depending both of fast and slow variables were studied in
\cite{BG}--\cite{B1}. In \cite{BG} the Schr\"odinger operator
with fast oscillating compactly supported potential was
considered. The object of the study was the phenomenon of new
eigenvalue emerging from the threshold of the continuous
spectrum. The paper \cite{B3} deals with a periodic operator
(independent of the small parameter) perturbed by a fast
oscillating compactly supported potential with increasing
amplitude. Here the structure and the behavior of the spectrum
were studied in details.  In \cite{B1} they studied the
Schr\"odinger operator with a compactly supported potential
independent of the small parameter; the perturbation was a fast
oscillating periodic potential. The asymptotic behavior of the
spectrum was described. We note that the homogenization of the
resolvent was not considered in \cite{BG}--\cite{B1}. At the
same time, the technique employed is sufficient to study this
question and to obtain the results analogous to \cite{BS}.

In the present paper we consider a quite general second order
elliptic system in the whole space. The first difference to the
operators considered in \cite{Bi}--\cite{BS5} is the presence of
the lower order terms. More precisely, the second order part of
our operator is written in the divergence form similar to the
cited papers. The lower terms are introduced quite arbitrarily;
the only restriction is that the operator is self-adjoint and
lower-semibounded uniformly in the small parameter.  The certain
smoothness for the coefficients is also assumed. One more
difference to the \cite{Bi}--\cite{Zh3} is that in our case the
coefficients depend both on slow and fast variables. The
dependence of fast variables is periodic. The coefficients are
supposed to be uniformly bounded w.r.t. to slow variables; the
same is supposed for certain derivatives of the coefficients.

In the work we construct the homogenized operator and obtain the
first terms of the asymptotic expansion for the resolvent of the
perturbed operator for all values of the spectral parameter
separated apriori from the spectrum of the homogenized operator.
These asymptotics are obtained for the resolvent treated as an
operator in $L_2$ as well as an operator from $L_2$ into $\H^1$.
We borrow the main ideas of \cite{Zh3} to obtain these results.
Moreover, we assume the coefficients to be smoother than in
\cite{Bi}--\cite{Zh3} that allows us to simplify certain details
in the arguments. In particular, it allows us to avoid the
smoothing used in the cited papers. In the end we give examples
of some operators to which our results can be applied.

\section{Formulation of the problem and the main results
}

Let $x=(x_1,\ldots,x_d)$ be the Cartesian coordinates in
$\mathbb{R}^d$, $d\geqslant 1$, $B=B(\z)$ be a matrix-valued
function,
\begin{equation*}
B(\z)=\sum\limits_{i=1}^{d}B_i\z_i,
\end{equation*}
where $\z=(\z_1,\ldots,\z_d)$, $B_i$ are constant complex-valued
matrices of the size $m\times n$, and $m\geqslant n$. Hereafter
we assume that $\rank B(\z)=n$, $\z\not=0$.

Let $Y$ be a Banach space. By $\Hinf^k(\mathbb{R}^d;Y)$ we
denote the Sobolev space of the functions defined on
$\mathbb{R}^d$ and having values in $Y$, and so that
\begin{equation*}
\|\mathbf{u}\|_{\Hinf^k(\mathbb{R}^d;Y)}:=\max
\limits_{|\a|\leqslant k}\,\esssup\limits_{x\in \mathbb{R}^d}
\Big\|\frac{\p^{|\a|} \mathbf{u}}{\p x^a}\Big\|_{Y}<\infty.
\end{equation*}
If $k=0$, we will employ the notation
$L_\infty(\mathbb{R}^d;Y)$.

In the space $\mathbb{R}^d$ we select a lattice; its elementary
cell is indicated by $\square$. We will employ the symbol
$C_{per}^\g(\overline{\square})$ to denote the space of
$\square$-periodic functions having finite H\"older norm
$\|\cdot\|_{C^\g(\overline{\square})}$. The norm in this space
coincides with the norm of $C^\g(\overline{\square})$.

We will often treat a $\square$-periodic w.r.t. $\xi$
vector-function $\mathbf{f}=\mathbf{f}(x,\xi)$ as mapping points
$x\in\mathbb{R}^d$ into the function depending of $\xi$. The map
is defined as $x\mapsto \mathbf{f}(x,\cdot)$. It will allow us
to speak about the belonging of the function $\mathbf{f}(x,\xi)$
to the spaces $\H^k(Q;C^\g_{per}(\overline{\square}))$ and
$\Hinf^k(Q;C^\g_{per}(\overline{\square}))$.

Let $A=A(x,\xi)$ be a matrix-valued function of the size
$m\times m$. We suppose that the matrix $A$ is hermitian and
$\square$-periodic w.r.t. $\xi$, and the uniform in  $(x,\xi)\in
\mathbb{R}^{2d}$ estimate
\begin{equation}\label{1.4}
c_1 E_m\leqslant A(x,\xi)\leqslant c_2 E_m,
\end{equation}
is valid, where $E_m$ is $m\times m$ unit matrix. We also assume
that $A\in
\Hinf^1(\mathbb{R}^d;C^{1+\b}_{per}(\overline{\square}))\cap
\Hinf^2(\mathbb{R}^d;C^{\b}_{per}(\overline{\square}))$ for some
$\b\in(0,1)$. By $V=V(x,\xi)$, $a_i=a_i(x,\xi)$ we denote
$\square$-periodic w.r.t. $\xi$ matrix-valued functions of the
size $n\times n$. It is assumed that $a_i\in
\Hinf^1(\mathbb{R}^d;C^{1+\b}_{per}(\overline{\square}))\cap
\Hinf^2(\mathbb{R}^d;C^{\b}_{per}(\overline{\square}))$, $V\in
\Hinf^1(\mathbb{R}^d;C^{\b}_{per}(\overline{\square}))$. It is
also supposed that the matrix $V$ is hermitian, and the matrices
$a_j$ and $B_j$ are complex-valued. Let $b_i=b_i(x)\in
\Hinf^2(\mathbb{R}^d)$ be complex-matrix-valued functions of the
size $n\times n$.

By $\e$ we denote a small positive parameter. Given a function
$f(x,\xi)$, by $f_\e(x)$ we indicate
$f\left(x,\frac{x}{\e}\right)$; for instance,
$A_\e(x):=A\left(x,\frac{x}{\e}\right)$.

The aim of the present work is to study the spectral properties
of the operator
\begin{align}
&\mathcal{H}_\e:=B(\p)^*A_\e B(\p)+a_\e(x,\p)+V_\e,\label{1.5}
\\
&a_\e(x,\p):=a\left(x,\frac{x}{\e},\p\right),\quad a(x,\xi,\z):=
\sum\limits_{i=1}^{d} \left(a_i(x,\xi)\z_{i}b_i(x)-
b_i^*(x)\z_{i}a_i^*(x,\xi)\right), \nonumber
\end{align}
in $L_2(\mathbb{R}^d;\mathbb{C}^n)$ with
$\H^2(\mathbb{R}^d;\mathbb{C}^n)$ as the domain. Here
$\p=(\p_1,\ldots,\p_d)$, $\p_i$ is the derivative w.r.t. $x_i$,
the superscript $^*$ indicated conjugation, and
\begin{equation*}
B(\p)^*:=-\sum\limits_{i=1}^{d} B_i^*\p_i. 
\end{equation*}
We will show that the operator $\mathcal{H}_\e$ is self-adjoint
and lower-semibounded uniformly in $\e$ (see Lemma~\ref{lm2.3}).

Let $\L_0=\L_0(x,\xi),\L_1=\L_1(x,\xi)$, $i=0,1$ be the matrices
of the size $n\times n$ and $n\times m$, respectively, being
$\square$-periodic w.r.t. $\xi$ solutions of the equations
\begin{equation}\label{1.10}
\begin{aligned}
&B(\p_\xi)^*A(x,\xi)B(\p_\xi)\L_0(x,\xi)-\sum\limits_{i=1}^{d}
b_i^*(x)\frac{\p a_i^*}{\p\xi_i}(x,\xi)=0,&& 
(x,\xi)\in\mathbb{R}^{2d},
\\
&B(\p_\xi)^*A(x,\xi)\big(B(\p_\xi)\L_1(x,\xi)+E_m\big)=0,&&
(x,\xi)\in\mathbb{R}^{2d},
\end{aligned}
\end{equation}
and satisfying the conditions
\begin{equation}\label{1.11}
\int\limits_\square \L_i(x,\xi)\di\xi=0,\quad x\in\mathbb{R}^d.
\end{equation}
Here
$\p_\xi=\left(\frac{\p}{\p\xi_1},\ldots,\frac{\p}{\p\xi_d}\right)$.
We will show below that the solutions to (\ref{1.10}),
(\ref{1.11}) exist, are unique, and  $\L_i\in
\Hinf^1(\mathbb{R}^d;C^{2+\b}_{per}(\overline{\square}))$ (see
the proof of Lemma~\ref{lm2.4}).

Let $\mathcal{H}_0$ be an operator in
$L_2(\mathbb{R}^d;\mathbb{C}^n)$ defined as
\begin{align}
&\mathcal{H}_0=B(\p)^*A_2 B(\p)+A_1(x,\p)+A_0, \label{1.13}
\\
&
\begin{aligned}
&A_2(x):=\frac{1}{|\square|}\int\limits_\square
A(x,\xi)\big(B(\p_\xi)\L_1(x,\xi)+E_m\big)\di\xi,
\\
&A_1(x,\p):=\frac{1}{|\square|}B(\p)^*\int\limits_\square
A(x,\xi) B(\p_\xi)\L_0(x,\xi)\di\xi
\\
&\hphantom{A_1(x,\p):=}+\left(\frac{1}{|\square|}\int\limits_\square
\big(B(\p_\xi)\L_0(x,\xi)\big)^* A(x,\xi)\di\xi\right)
B(\p)+\frac{1}{|\square|}\int\limits_\square a(x,\xi,\p)\di\xi,
\\
&A_0(x):=-\frac{1}{|\square|}\int\limits_\square
\big(B(\p_\xi)\L_0(x,\xi)\big)^*A(x,\xi)B(\p_\xi)\L_0(x,\xi)\di\xi
+\frac{1}{|\square|}\int\limits_\square V(x,\xi)\di\xi,
\end{aligned}\label{1.14}
\end{align}
on the domain $\H^2(\mathbb{R}^d;\mathbb{C}^n)$. We will show
below (see Lemma~\ref{lm2.4}) that this operator is self-adjoint
and lower-semibounded, and its coefficients are smooth enough
(see Lemma~\ref{lm2.4}). By  $\mathfrak{h}_0$ we denote the
lower bound of $\mathcal{H}_0$.

Let $G=G(x,\xi)\in
\Hinf^1(\mathbb{R}^d;C^{\b}_{per}(\overline{\square}))$  be a
positive hermitian matrix of the size $n\times n$. We also
assume that the inverse matrix $G^{-1}$ is uniformly bounded. By
$\mathfrak{g}_i>0$, $i=1,2$, we denote constants independent of
$\e$, $x$ and $\xi$ such that
\begin{equation}\label{1.6a}
\mathfrak{g}_1 E_n\leqslant G(x,\xi) \leqslant \mathfrak{g}_2
E_n.
\end{equation}
We let
\begin{equation*}
G_0(x):=\frac{1}{|\square|}\int\limits_{\square} G(x,\xi)\di\xi.
\end{equation*}

We introduce an operator
\begin{equation}\label{1.7a}
\mathcal{L}_\e:=\left(\L_1\left(x,\frac{x}{\e}\right)B(\p)+
\L_0\left(x,\frac{x}{\e}\right)\right).
\end{equation}
It will be shown that for each $\e$ the operator
$\mathcal{L}_\e$ is bounded as one from
$\H^1(\mathbb{R}^d;\mathbb{C}^n)$ into $
L_2(\mathbb{R}^d;\mathbb{C}^n)$ and from
$\H^2(\mathbb{R}^d;\mathbb{C}^n)$ into
$\H^1(\mathbb{R}^d;\mathbb{C}^n)$ (see Lemma~\ref{lm3.5}).

Our first result describes the approximation for the generalized
resolvent of $\mathcal{H}_\e$.

\begin{theorem}\label{th1.2}
Suppose $\l\in\mathbb{C}\setminus[\mu_0,+\infty)$,
$\mu_0:=\min\left\{\frac{\mathfrak{h}_0}{\mathfrak{g}_1},
\frac{\mathfrak{h}_0}{\mathfrak{g}_2}\right\}$. Then for all
small $\e$ the inequalities
\begin{equation}\label{1.15}
\begin{aligned}
&\|(\mathcal{H}_\e-\l G_\e)^{-1}-(\mathcal{H}_0-\l G_0
)^{-1}\|_{L_2\to L_2}\leqslant C\e,
\\
&\|(\mathcal{H}_\e-\l G_\e)^{-1}- (\I+\e
\mathcal{L}_\e)(\mathcal{H}_0-\l G_0)^{-1}\|_{L_2\to
\H^1}\leqslant C\e,
\end{aligned}
\end{equation}
hold true, where $\I$ is the identical operator, the constants
$C$ are independent of $\e$, and  and the norms are regarded as
those for the operators from $L_2(\mathbb{R}^d;\mathbb{C}^n)$
into $L_2(\mathbb{R}^d;\mathbb{C}^n)$ and
$\H^1(\mathbb{R}^d;\mathbb{C}^n)$, respectively.
\end{theorem}

Hereafter by $\spec(\cdot)$ we denote the spectrum.

\begin{corollary}\label{th1.1}
The spectrum of $\mathcal{H}_\e$ converges to the spectrum of
$\mathcal{H}_0$. Namely, if $\l\not\in\spec(\mathcal{H}_0)$, it
follows that $\l\not\in\spec(\mathcal{H}_\e)$ for all $\e$ small
enough, and if $\l\in\spec(\mathcal{H}_0)$, there exists
$\l_\e\in\spec(\mathcal{H}_\e)$ such that $\l_\e\to\l_0$ as
$\e\to+0$. If $\a_1,\a_2\in \mathbb{R}\setminus
\spec(\mathcal{H}_0)$, then the spectral projectors of
$\mathcal{H}_\e$ and $\mathcal{H}_0$ satisfy the convergence
сходимость $\mathcal{P}_{(\a_1,\a_2)}(\mathcal{H}_\e)\to
\mathcal{P}_{(\a_1,\a_2)}(\mathcal{H}_0)$, $\e\to+0$.
\end{corollary}

We should say that in the papers \cite{BS2}, \cite{BS5},
\cite{Zh3} they considered a particular case of the operator
$\mathcal{H}_\e$ corresponding to the identities $a_i=0$,
$b_i=0$, $V=0$, and also under the assumption $A=A(\xi)$. In
this case the estimates similar to (\ref{1.15}) were obtained.
It should be stressed that in the cited papers the matrices $A$
and $G$ were not assumed to be smooth, but bounded only.
Moreover, the constants  $C$ in the mentioned estimates depended
only of $L_\infty$-norm of the matrices $A$, $G$, $G^{-1}$ as
well as of the lattice. In our case these constants depend of
$\l$, the lattice, and the norms of the coefficients in the
spaces the belong to.

\section{Auxiliary statements}

In the present section we prove a series of auxiliary statements
required for the proofs of Theorem~\ref{th1.2} and
Corollary~\ref{th1.1}.

\begin{lemma}\label{lm2.1}
For any $\mathbf{u}\in\H^1(\mathbb{R}^d;\mathbb{C}^n)$ the
uniform in $\e$ estimate
\begin{equation*} C_1\|\nabla \mathbf{u}
\|_{L_2(\mathbb{R}^d;\mathbb{C}^n)}^2 \leqslant \big(A_\e
B(\p)\mathbf{u},B(\p)\mathbf{u}\big)_{L_2(\mathbb{R}^d;\mathbb{C}^m)}
\leqslant C_2 \|\nabla\mathbf{u}
\|_{L_2(\mathbb{R}^d;\mathbb{C}^n)}^2. 
\end{equation*}
holds true.
\end{lemma}

\begin{proof}
It follows from (\ref{1.4}) that
\begin{equation*}
c_1\|B(\p)\mathbf{u}\|_{L_2(\mathbb{R}^d;\mathbb{C}^n)}^2\leqslant
\big(A_\e
B(\p)\mathbf{u},B(\p)\mathbf{u}\big)_{L_2(\mathbb{R}^d;\mathbb{C}^m)}
\leqslant
c_2\|B(\p)\mathbf{u}\|_{L_2(\mathbb{R}^d;\mathbb{C}^n)}^2.
\end{equation*}
The desired inequality follows now from \cite[Ch. 2, \S 1,
estimate (1.11)]{BS}.
\end{proof}

\begin{lemma}\label{lm2.3}
The operator $\mathcal{H}_\e$ is self-adjoint and
lower-semibounded uniformly in $\e$.
\end{lemma}

\begin{proof}
The semiboundedness follows easily from the properties of the
coefficients of   $\mathcal{H}_\e$ and the identity
\begin{align*}
(\mathcal{H}_\e
\mathbf{u},\mathbf{u})_{L_2(\mathbb{R}^d;\mathbb{C}^n)}&=
h_\e[\mathbf{u}]:=\big(A_\e B(\p)\mathbf{u},
B(\p)\mathbf{u}\big)_{L_2(\mathbb{R}^d;\mathbb{C}^m)}
\\
&+ 2\RE\sum\limits_{i=1}^{d}\left(a_{i,\e} \p_{i}b_i \mathbf{u},
\mathbf{u}\right)_{L_2(\mathbb{R}^d;\mathbb{C}^n)} +
\big(V_\e\mathbf{u},
\mathbf{u}\big)_{L_2(\mathbb{R}^d;\mathbb{C}^n)}.
\end{align*}

It is clear that the operator $\mathcal{H}_\e$ is symmetric; to
prove the self-adjointness it is sufficient to check that
$\mathcal{D}(\mathcal{H}_\e^*)=\mathcal{D}(\mathcal{H}_\e)$. In
its turn, this identity can be established easily, if a
generalized solution to the equation
\begin{equation}\label{2.1a}
\left(B(\p)^*A_\e B(\p)+a_\e(x,\p)+V_\e\right)\mathbf{u}=
\mathbf{f},\quad x\in\mathbb{R},\qquad \mathbf{f}\in
L_2(\mathbb{R}^d;\mathbb{C}^n),
\end{equation}
belongs to $\H^2(\mathbb{R}^d;\mathbb{C}^n)$. Let us prove it.

It is clear that a generalized solution of  (\ref{2.1a}) is also
a generalized solution to
\begin{gather}
B(\p)^*A_\e B(\p)\mathbf{u}+\mathbf{u}=\mathbf{g},\quad
x\in\mathbb{R}^d, \label{2.2a}
\\
\mathbf{g}:=\mathbf{f}-a_\e(x,\p)\mathbf{u}-
V_\e\mathbf{u}+\mathbf{u},\quad
\|\mathbf{g}\|_{L_2(\mathbb{R}^d;\mathbb{C}^n)}\leqslant C
\left(\|\mathbf{f}\|_{L_2(\mathbb{R}^d;\mathbb{C}^n)}+
\|\mathbf{u}\|_{\H^1(\mathbb{R}^d;\mathbb{C}^n)} \right).
\nonumber
\end{gather}
Let $\d\not=0$ be a small fixed number, $\mathbf{e}_i^{(d)}$,
$i=1,\ldots,d$ be a standard basis in $\mathbb{R}^d$. We denote
\begin{equation*}
\mathbf{u}_\d^{(i)}(x):=\frac{1}{\d}\left(\mathbf{u}(x+\d
\mathbf{e}_i^{(d)}) -\mathbf{u}(x)\right).
\end{equation*}
This function is a generalized solution to  (\ref{2.2a}) with
the right hand side
\begin{equation*}
\mathbf{g}_\d^{(i)}-B(\p)^*A_{\e,\d}^{(i)} B(\p)
\mathbf{u}(x+\d\mathbf{e}_i^{(d)}),
\end{equation*}
where $\mathbf{g}_\d^{(i)}$, $A_{\e,\d}^{(i)}$ are defined via
$\mathbf{g}$, $A_\e$ similarly to $\mathbf{u}_\d^{(i)}$. The
integral identity corresponding to the equation for
$\mathbf{u}_\d^{(i)}$ reads as follows
\begin{align*}
&\left(A_\e B(\p)\mathbf{u}_\d^{(i)},
B(\p)\boldsymbol{\vp}\right)_{L_2(\mathbb{R}^d;\mathbb{C}^m)}+
\left(\mathbf{u}_\d^{(i)},
\boldsymbol{\vp}\right)_{L_2(\mathbb{R}^d;\mathbb{C}^n)}
\\
&=- \left(\mathbf{g},\boldsymbol{\vp}_{-\d}^{(i)}
\right)_{L_2(\mathbb{R}^d;\mathbb{C}^n)}- \left(A_{\e,\d}^{(i)}
B(\p)\mathbf{u}(\cdot+\d\mathbf{e}_i^{(d)}),
B(\p)\boldsymbol{\vp}\right)_{L_2(\mathbb{R}^d;\mathbb{C}^n)},
\end{align*}
where $\boldsymbol{\vp}\in\H^1(\mathbb{R}^d;\mathbb{C}^n)$. We
also observe that inequality (11) in the proof of Item~a) of
Theorem~3 in \cite[Ch. I\!I\!I, Sec. 3.4]{Mi} implies
\begin{equation*}
\|\boldsymbol{\vp}_{-\d}^{(i)}\|_{L_2(\mathbb{R}^d;\mathbb{C}^n)}
\leqslant
\|\boldsymbol{\vp}\|_{\H^1(\mathbb{R}^d;\mathbb{C}^n)},
\end{equation*}
for each $\boldsymbol{\vp}\in\H^1(\mathbb{R}^d;\mathbb{C}^n)$.
Letting $\boldsymbol{\vp}:=\mathbf{u}_\d^{(i)}$ in two last
inequalities and taking into account the smoothness of $A$ and
Lemma~\ref{lm2.1}, we arrive at the uniform in $\d$ estimate
\begin{equation*}
\|\mathbf{u}_\d^{(i)}\|_{\H^1(\mathbb{R}^d;\mathbb{C}^n)}
\leqslant
C\left(\|\mathbf{g}\|_{L_2(\mathbb{R}^d;\mathbb{C}^n)}+
\|\mathbf{u}\|_{\H^1(\mathbb{R}^d;\mathbb{C}^n)} \right).
\end{equation*}
Employing this estimate and repeating the arguments of the proof
of Item~b) of Theorem~3 in \cite[Гл. I\!I\!I, \S 3.4]{Mi}, one
can check easily that there exist second generalized derivations
of the function $\mathbf{u}$, and
\begin{equation}\label{2.2c}
\|\mathbf{u}\|_{\H^2(\mathbb{R}^d;\mathbb{C}^n)}\leqslant C
\left(\|\mathbf{g}\|_{L_2(\mathbb{R}^d;\mathbb{C}^n)} +
\|\mathbf{u}\|_{\H^1(\mathbb{R}^d;\mathbb{C}^n)}\right).
\end{equation}
\end{proof}

\begin{lemma}\label{lm3.0}
Let $\mathbf{f}(x,\cdot)\in C_{per}^\b(\overline{\square})$ for
all $x\in\mathbb{R}^d$. The system
\begin{equation}\label{3.6}
B(\p_\xi)^*A(x,\xi)B(\p_\xi)
\mathbf{v}(x,\xi)=\mathbf{f}(x,\xi),\quad \xi\in\mathbb{R}^d,
\end{equation}
has $\square$-periodic w.r.t. $\xi$ solution
$\mathbf{v}(x,\cdot)\in C_{per}^{2+\b}(\overline{\square})$, if
and only if
\begin{equation}\label{3.7}
\int\limits_\square \mathbf{f}(x,\xi)\di \xi=0,\quad
x\in\mathbb{R}^d.
\end{equation}
If the solvability condition holds true, the solution of
(\ref{3.6}) is unique up to a constant (in $\xi$) vector. There
exists unique solution of (\ref{3.6}) such that
\begin{equation}\label{3.8}
\int\limits_\square \mathbf{v}(x,\xi)\di\xi=0,\quad
x\in\mathbb{R}^d.
\end{equation}
This solution satisfies the
estimate
\begin{equation}\label{2.10a}
\|\mathbf{v}(x,\cdot)\|_{C_{per}^{2+\b}(\overline{\square})}
\leqslant
C\|\mathbf{f}(x,\cdot)\|_{C_{per}^\b(\overline{\square})},
\end{equation}
where the constant $C$ is independent of $x\in\mathbb{R}^d$ and
$\mathbf{f}$. If $\mathbf{f}\in
\Hinf^k\big(\mathbb{R}^d;C_{per}^\b (\overline{\square})\big)$,
$k=0,1$ it follows that $\mathbf{v}\in
\Hinf^k\big(\mathbb{R}^d;C_{per}^{2+\b}
(\overline{\square})\big)$, and the estimate
\begin{equation}
\|\mathbf{v}\|_{\Hinf^k(\mathbb{R}^d;C_{per}^{2+\b}(\overline{\square}))}\leqslant
C\|\mathbf{f}\|_{\Hinf^k(\mathbb{R}^d;C_{per}^\b(\overline{\square}))},
\label{2.11}
\end{equation}
holds true, where the constant $C$ is independent of
$\mathbf{f}$.
\end{lemma}

\begin{proof}
The existence of a $\square$-periodic generalized solution of
(\ref{3.6}) in $\H^1(\square;\mathbb{C}^n)$, the solvability
condition (\ref{3.7}) and the uniqueness of the solution
satisfying (\ref{3.8}) are implied by Theorem~1 in
\cite[Appendix]{BP}. Moreover, it follows from the proof of this
theorem that
\begin{equation}\label{2.11a}
\|\mathbf{v}(x,\cdot)\|_{\H^1(\square;\mathbb{C}^n)}\leqslant
C\|\mathbf{f}(x,\cdot)\|_{L_2(\square;\mathbb{C}^n)}.
\end{equation}
Throughout the proof by $C$ we indicate the inessential
constants independent of  $\mathbf{f}$ and $x\in \mathbb{R}^d$.

As for the equation (\ref{2.1a}), one can make sure that
$\mathbf{v}(x,\cdot)\in\Hloc^2(\mathbb{R}^d)$. By theorem~10.7
and Remark~1 in \cite[Ch. I\! V, \S 10.3]{ADN1} and the
periodicity of $\mathbf{f}$ and $\mathbf{v}$ it implies that
\begin{equation*}
\|\mathbf{v}(x,\cdot)\|_{C^{2+\b}(\overline{\square})}\leqslant
C\big(\|\mathbf{f}(x,\cdot)\|_{C^\b(\overline{\square})}+
\|\mathbf{v}(x,\cdot)\|_{L_1(\square)}\big).
\end{equation*}
This estimate and (\ref{2.11a}) yield (\ref{2.10a}).

Assume that $\mathbf{f}\in
\Hinf^k(\mathbb{R}^d;C_{per}^{2+\b}(\overline{\square}))$ and
let us prove the claimed smoothness of $\mathbf{v}$. If $k=0$
and $\mathbf{f}\in
L_{\infty}\big(\mathbb{R}^d;C_{per}^{2+\b}(\overline{\square})\big)$,
the belonging $\mathbf{v}\in
L_{\infty}\big(\mathbb{R}^d;C_{per}^{2+\b}(\overline{\square})\big)$
and the estimate (\ref{2.11}) follow immediately from
(\ref{2.10a}). Let $k=1$. We take a small number $\d\not=0$,
choose a point $x\in{Q}$, and indicate
\begin{equation*}
\mathbf{v}_\d^{(i)}(x,\xi):=\frac{1}{\d}
\big(\mathbf{v}(x+\d\mathbf{e}_i^{(d)},\xi)-\mathbf{v}(x,\xi)\big). 
\end{equation*}
This function is the solution to (\ref{3.6}) at $x$ with the
right hand side
\begin{equation*}
\mathbf{f}_\d^{(i)}-B(\p_\xi)A_{\e,\d}^{(i)}
B(\p_\xi)\mathbf{v}(x+\d\mathbf{e}_i^{(d)},\xi),
\end{equation*}
where $\mathbf{f}_\d^{(i)}$ and $A_{\e,\d}^{(i)}$ are determined
by analogy with $\mathbf{v}_\d^{(i)}$. This right hand side
satisfies (\ref{3.7}) and belongs to
$L_{\infty}(\mathbb{R}^d;C_{per}^\b(\overline{\square}))$. It
follows now from (\ref{2.11}) for $k=0$ that
\begin{equation}\label{2.9a}
\begin{aligned}
\|\mathbf{v}_\d^{(i)}\|_{L_{\infty}(\mathbb{R}^d;
C^{2+\b}_{per}(\overline{\square}))}\leqslant & C
\|\mathbf{f}_\d^{(i)}-B(\p_\xi)A_{\e,\d}^{(i)}
B(\p_\xi)\mathbf{v}(\cdot+\d,\cdot)\|_{L_{\infty}(\mathbb{R}^d;
C_{per}^\b(\overline{\square}))}
\\
\leqslant & C\|\mathbf{f}\|_{C_1^*(\mathbb{R}^d;
C_{per}^\b(\overline{\square}))},
\end{aligned}
\end{equation}
where the constant $C$ is independent of $\mathbf{f}$ and $\d$.

Let $\mathbf{v}_0^{(i)}$ be a solution to (\ref{3.6}) at $x$
with the right-hand side
\begin{equation*} 
\frac{\p\mathbf{f}}{\p x_i}(x,\xi)- B(\p_\xi)^*\frac{\p A}{\p
x_i}(x,\xi) B(\p_\xi) \mathbf{v},
\end{equation*}
satisfying (\ref{3.8}). It is clear that the right hand side
satisfies (\ref{3.7}) and belongs to
$L_{\infty}(\mathbb{R}^d;C_{per}^{\b}(\overline{\square}))$. The
function $\mathbf{v}_\d^{(i)}-\mathbf{v}_0^{(i)}$ is the
solution to (\ref{3.6}) at $x$ with the right hand side
\begin{equation*} 
\mathbf{f}_\d^{(i)}-\frac{\p \mathbf{f}}{\p x_i}+B(\p_\xi)^*
\left(\frac{\p A}{\p x_i
}-A_\d^{(i)}\right)B(\p_\xi)\mathbf{v}-\d
B(\p_\xi)^*A_\d^{(i)}B(\p_\xi)\mathbf{v}_\d^{(i)}.
\end{equation*}
Employing now (\ref{2.10a}) and (\ref{2.9a}), we obtain
\begin{equation*}
\|\mathbf{v}_\d^{(i)}-
\mathbf{v}_0^{(i)}\|_{L_{\infty}(\mathbb{R}^d;
C_{per}^{2+\b}(\overline{\square})} \xrightarrow[\d\to0]{}0.
\end{equation*}
Hence, the derivative $\frac{\p\mathbf{v}}{\p x_i}$ 
exists and $\frac{\p\mathbf{v}}{\p x_i} =\mathbf{v}_0^{(i)}\in
L_{\infty}(\mathbb{R}^d;C^{2+\b}_{per}(\overline{\square}))$.
The inequality (\ref{2.10a}) implies also that the inequality
(\ref{2.11}) is valid with $k=1$.
\end{proof}

\begin{lemma}\label{lm2.4}
The operator $\mathcal{H}_0$ is self-adjoint and lower
semibounded. The matrices $A_2$, $A_0$, and the coefficients of
$A_1(x,\p)$ belong to $\Hinf^1(\mathbb{R}^d)$. The estimate
\begin{equation}\label{2.11b}
\|\mathbf{u}\|_{\H^2(\mathbb{R}^d;\mathbb{C}^n)}\leqslant
C\left(\|\mathcal{H}_0\mathbf{u}\|_{L_2(\mathbb{R}^d;\mathbb{C}^n)}
+\|\mathbf{u}\|_{L_2(\mathbb{R}^d;\mathbb{C}^n)}\right)
\end{equation}
is valid.
\end{lemma}

\begin{proof}
We begin with the proof of the solvability of  (\ref{1.10}),
(\ref{1.11}). Let
$\boldsymbol{\L}_0^{(j)}=\boldsymbol{\L}_0^{(j)}(x,\xi)\in
\mathbb{C}^n$, $j=1,\ldots,n$,
$\boldsymbol{\L}_1^{(j)}=\boldsymbol{\L}_1^{(j)}(x,\xi)\in
\mathbb{C}^n$, $j=1,\ldots,m$, be $\square$-periodic w.r.t.
$\xi$ solutions of
\begin{align*}
&B(\p_\xi)^*A B(\p_\xi)\boldsymbol{\L}_0^{(j)}- 
\sum\limits_{i=1}^{d} b_i^*\frac{\p a_i^*}{\p\xi_i}
\mathbf{e}_j^{(n)}=0,&&(x,\xi)\in\mathbb{R}^{2d},
\\
&B(\p_\xi)^*A\big(B(\p_\xi)\boldsymbol{\L}_1^{(j)}+
\mathbf{e}_j^{(m)}\big)=0,&& (x,\xi)\in\mathbb{R}^{2d},
\end{align*}
satisfying (\ref{3.8}). Here $A=A(x,\xi)$, $a_i=a_i(x,\xi)$,
$\mathbf{e}_j^{(n)}$, $j=1,\ldots,n$ is the standard basis in
$\mathbb{C}^n$.  Lemma~\ref{lm3.0} implies that these problems
are uniquely solvable and their solutions belong to
$\Hinf^1(\mathbb{R}^d;C_{per}^{2+\b}(\overline{\square}))$. The
vectors $\boldsymbol{\L}_i^{(j)}$ are columns of the matrices
$\L_i$. Hence, the matrices $A_2$, $A_0$, as well as the
coefficients of the operator $A_1(x,\p)$ belong to
$\Hinf^1(\mathbb{R}^d;\mathbb{C}^n)$.

We denote $X:=\big(B(\p_\xi)\L_1+E_m\big)$. Integrating by parts
and taking into account the equations (\ref{1.10}) we obtain
\begin{align*}
\int\limits_{\square} &\big(B(\p_\xi)\L_1(x,\xi)\big)^*
A(x,\xi)X(x,\xi)\di\xi
\\
&=\int\limits_\square \L_1^*(x,\xi)B(\p_\xi)^*
A(x,\xi)\big(B(\p_\xi)\L_1(x,\xi)+E_m\big)\di\xi=0,
\end{align*}
that implies
\begin{equation}\label{3.12}
A_2(x)=\frac{1}{|\square|}\int\limits_{\square} X^*(x,\xi)
A(x,\xi) X(x,\xi)\di\xi.
\end{equation}
This identity and the definition of lower terms of
$\mathcal{H}_0$ yield that this operator is symmetric. Let us
prove that it is lower-semibounded. In view of the last identity
and  (\ref{1.4}) it easy to check that for all
$x\in\mathbb{R}^d$ and $\mathbf{w}\in \mathbb{C}^n$ the
inequalities
\begin{align*}
&(A_2(x)\mathbf{w},\mathbf{w})_{\mathbb{C}^m}=\Big(A(x,\cdot)
X(x,\cdot)\mathbf{w}, X(x,\cdot)\mathbf{w}
\Big)_{L_2(\square;\mathbb{C}^m)}
\\
&\geqslant c_1 \left(\|\mathbf{w}\|_{\mathbb{C}^n}^2|\square|+
2\RE\big(B(\p_\xi)\L_1(x,\cdot)\mathbf{w},
\mathbf{w}\big)_{L_2(\square;\mathbb{C}^n)}+
\|B(\p_\xi)\L_1(x,\cdot)
\mathbf{w}\|_{L_2(\square;\mathbb{C}^n)}^2\right)
\\
&=c_1 \left(\|\mathbf{w}\|_{\mathbb{C}^n}^2|\square|+
\|B(\p_\xi)\L_1(x,\cdot)
\mathbf{w}\|_{L_2(\square;\mathbb{C}^n)}^2\right)\geqslant
c_1|\square|\|\mathbf{w}\|_{\mathbb{C}^n}^2
\end{align*}
are valid. Together with the inequality (1.11) in \cite[Ch. 2,
\S 1]{BS} they imply
\begin{equation*}
\left\|\nabla\mathbf{u} 
\right\|_{L_2(\mathbb{R}^d;\mathbb{C}^n)}^2 \leqslant C
\left(A_2 B(\p)\mathbf{u},
B(\p)\mathbf{u}\right)_{L_2(\mathbb{R}^d;\mathbb{C}^m)}.
\end{equation*}
Employing this estimate by analogy with~\ref{lm2.3} one can
check easily that $\mathcal{H}_0$ is self-adjoint. The estimate
(\ref{2.11b}) follows easily from the corresponding analogue of
(\ref{2.2c}) and an obvious estimate
\begin{equation*}
\|\mathbf{u}\|_{\H^1(\mathbb{R}^d;\mathbb{C}^n)}\leqslant C
\left(\|\mathcal{H}_0\mathbf{u}\|_{L_2(\mathbb{R}^d;\mathbb{C}^n)}
+\|\mathbf{u}\|_{L_2(\mathbb{R}^d;\mathbb{C}^n)} \right).
\end{equation*}
\end{proof}

\section{Asymptotics for the resolvent}

In the section we prove Theorem~\ref{th1.2} and
Corollary~\ref{th1.1}. For the proofs, we require four lemmas.

\begin{lemma}\label{lm3.5}
For each value $\e>0$ the operator $\mathcal{L}_\e$ defined by
(\ref{1.7a}) is bounded as an operator from
$\H^2(\mathbb{R}^d;\mathbb{C}^n)$ into
$\H^1(\mathbb{R}^d;\mathbb{C}^n)$. The operator $\mathcal{L}_\e$
is bounded uniformly in $\e$ as an operator from
$\H^1(\mathbb{R}^d;\mathbb{C}^n)$ into
$L_2(\mathbb{R}^d;\mathbb{C}^n)$.
\end{lemma}

The lemma follows from the belonging  $\L_i\in
\Hinf^1(\mathbb{R}^d;C^{2+\b}(\overline{\square}))$ proven in
Lemma~\ref{lm2.4}.

Hereinafter by $\frac{\p u}{\p x_i}$ we denote the partial
derivatives w.r.t. $x_i$ for the functions
$u=u\left(x,\frac{x}{\e}\right)$ treated as ones of independent
variables $x$ and $\xi=\frac{x}{\e}$. In the same way we regard
the partial derivatives $\frac{\p u}{\p\xi_i}$. We also remind
that we employ the symbols $\p_i$ to denote the full derivatives
w.r.t. $x_i$, i.e.,
\begin{equation}\label{3.0a}
\p_i u\left(x,\frac{x}{\e}\right)=\frac{\p u}{\p x_i}
\left(x,\frac{x}{\e}\right)+\e^{-1} \frac{\p u}{\p \xi_i}
\left(x,\frac{x}{\e}\right),
\end{equation}
and $\p=(\p_1,\ldots,\p_d)$.

\begin{lemma}\label{lm3.6}
Let $M=M(x,\xi)$ be a $\square$-periodic w.r.t. $\xi$ matrix of
the size $n\times n$ such that
\begin{equation}
M\in
\Hinf^1(\mathbb{R}^d;C_{per}^{\b}(\overline{\square})),\quad
\int\limits_{\square} M(x,\xi)\di\xi=0,\quad x\in\mathbb{R}^d,
\end{equation}
and $\mathbf{u}(x)\in \H^1(\mathbb{R}^d;\mathbb{C}^n)$ be a
vector-function. There exist vector-functions
$\mathbf{v}_i^{(\e)}(x)\in L_2(\mathbb{R}^d;\mathbb{C}^n)$,
$i=0,\ldots,d$, such that the representation and estimates
\begin{equation*}
M\left(x,\frac{x}{\e}\right)\mathbf{u}(x)=\e\sum\limits_{i=1}^{d}
\p_{i}\mathbf{v}_i^{(\e)}(x)+\e \mathbf{v}_0^{(\e)}(x),\quad
\|\mathbf{v}_i^{(\e)}\|_{L_2(\mathbb{R}^d;\mathbb{C}^n)}\leqslant
C\|\mathbf{u}\|_{\H^1(\mathbb{R}^d;\mathbb{C}^n)},
\end{equation*}
are valid, where the constant $C$ is independent of $\e$ and
$\mathbf{u}$.
\end{lemma}

\begin{proof}
Let $P=P(x,\xi)$ be a $\square$-periodic w.r.t. $\xi$ matrix of
the size $n\times n$ satisfying the equation
\begin{equation*}
\D_\xi P(x,\xi)=M(x,\xi),\quad (x,\xi)\in\mathbb{R}^{2d}, 
\end{equation*}
and the condition (\ref{3.8}) for all $x\in\mathbb{R}^d$. By
Lemma~\ref{lm3.0} this equation is solvable, the matrix $P$ is
defined uniquely, and the estimate $\|P\|_{\Hinf^1(\mathbb{R}^d;
C_{per}^{2+\b}(\overline{\square}))}<\infty$ is valid. Using
this estimate one can check easily that the lemma is valid for
\begin{equation*}
\mathbf{v}_i^{(\e)}:=\frac{\p P}{\p\xi_i}\mathbf{u},\quad
\mathbf{v}_0^{(\e)}:=-\sum\limits_{i=1}^{d}\frac{\p^2
P\mathbf{u}}{\p x_i\p\xi_i},\quad
P=P\left(x,\frac{x}{\e}\right).
\end{equation*}
\end{proof}

We denote
\begin{equation}\label{3.2a}
\widehat{A}_1(x,\xi):=A(x,\xi)(B(\p_\xi)\L_1(x,\xi)+E_m)-
A_2(x).
\end{equation}

\begin{lemma}\label{lm3.7}
Suppose $\mathbf{u}\in\H^2(\mathbb{R}^d;\mathbb{C}^n)$.  There
exist vector-functions $\mathbf{v}^{(\e)}_i\in
L_2(\mathbb{R}^d;\mathbb{C}^n)$, $i=1,\ldots,d$, such that the
representation and estimate
\begin{equation*}
B(\p_x)^*\widehat{A}_{1}\left(x,\frac{x}{\e}\right)
\mathbf{v}(x)=\e \sum\limits_{i=1}^{d}
\p_{i}\mathbf{v}_i^{(\e)}(x),\quad
\|\mathbf{v}^{(\e)}_i\|_{L_2(\mathbb{R}^d;\mathbb{C}^n)}\leqslant
C\|\mathbf{v}\|_{\H^1(\mathbb{R}^d;\mathbb{C}^n)},
\end{equation*}
are valid, where the constant $C$ is independent of $\e$ and
$\mathbf{u}$.
\end{lemma}

\begin{proof}
Let $P^{(i)}=P^{(i)}(x,\xi)$ be $\square$-periodic w.r.t. $\xi$
matrices of the size $n\times m$ satisfying equations
\begin{equation}\label{3.24}
\D_\xi P^{(i)}(x,\xi)=-B_i^*\widehat{A}_1(x,\xi),
\quad(x,\xi)\in\mathbb{R}^{2d},
\end{equation}
and the condition (\ref{3.8}). By Lemma~\ref{lm3.0} these
equations are solvable, the matrices $P^{(i)}$ are uniquely
defined, and $P^{(i)}\in
\Hinf^1(\mathbb{R}^d;C_{per}^{2+\b}(\overline{\square}))$. The
equations (\ref{1.10}) and the definition of $\widehat{A}$
implies that this matrix is $\square$-periodic w.r.t. $\xi$ and
satisfies the equation
\begin{equation}\label{3.23}
B(\p_\xi)^*\widehat{A}_1(x,\xi)=0,\quad
(x,\xi)\in\mathbb{R}^{2d}.
\end{equation}
This equation and (\ref{3.24}) yield that
\begin{equation}\label{3.4a}
\D_\xi\sum\limits_{i=1}^{d}\frac{\p P^{(i)}}{\p\xi_i}=0,\quad
(x,\xi)\in \mathbb{R}^d,
\end{equation}
and by the unique solvability of this equation we thus obtain
\begin{equation*}
\sum_{i=1}^{d}\frac{\p P^{(i)}}{\p\xi_i}=0.
\end{equation*}
Together with (\ref{3.23}) it follows that
\begin{equation}\label{3.4c}
-B_i^*\widehat{A}_1=\sum\limits_{j=1}^{d}\frac{\p
M_{ij}}{\p\xi_j}, \quad M_{ij}:=\frac{\p
P^{(i)}}{\p\xi_j}-\frac{\p P^{(j)}}{\p\xi_i}.
\end{equation}
Taking these identities into account we arrive at
\begin{equation}\label{3.4d}
B(\p_x)^*\widehat{A}_1\mathbf{v} = \sum\limits_{i,j=1}^{d}
\frac{\p^2 M_{ij}\mathbf{v}}{\p x_i\p\xi_j}=\e
\sum\limits_{i,j=1}^{d} \p_j\frac{\p M_{ij}\mathbf{v}}{\p
x_i}-\e\sum\limits_{i,j=1}^{d}\frac{\p^2 M_{ij} \mathbf{v}}{\p
x_i\p x_j},
\end{equation}
where $\widehat{A}_1=\widehat{A}_1\left(x,\frac{x}{\e}\right)$,
$M_{ij}=M_{ij}\left(x,\frac{x}{\e}\right)$. The second term in
the right hand side of the identity obtained is zero due to
$M_{ij}=-M_{ji}$. Now the statement of the lemma follows from
the belonging $P^{(i)}\in
\Hinf^1(\mathbb{R}^d;C_{per}^{2+\b}(\overline{\square}))$, if we
denote
\begin{equation*}
\mathbf{v}_i^{(\e)}(x):=\sum\limits_{j=1}^{d} \frac{\p}{\p x_j}
M_{ji}\left(x,\frac{x}{\e}\right)\mathbf{v}(x).
\end{equation*}
\end{proof}

We denote
\begin{equation}\label{3.4e}
\widehat{A}_0(x,\xi):=A(x,\xi)B(\p_\xi)\L_0(x,\xi)
-\frac{1}{|\square|} \int\limits_\square
A(x,\xi)B(\p_\xi)\L_0(x,\xi)\di\xi.
\end{equation}

\begin{lemma}\label{lm3.3a}
Let $\mathbf{v}\in\H^2(\mathbb{R}^d;\mathbb{C}^n)$. There exist
vector-functions $\mathbf{v}^{(\e)}_i\in
L_2(\mathbb{R}^d;\mathbb{C}^n)$, $i=0,\ldots,d$, such that the
representation and the estimate
\begin{equation*}
B(\p_x)^*\widehat{A}_{0}\left(x,\frac{x}{\e}\right)
\mathbf{v}(x)=\e \sum\limits_{i=1}^{d}
\p_{i}\mathbf{v}_i^{(\e)}(x)+\e \mathbf{v}_0^{(\e)}(x),\quad
\|\mathbf{v}^{(\e)}_i\|_{L_2(\mathbb{R}^d;\mathbb{C}^n)}\leqslant
C\|\mathbf{v}\|_{\H^2(\mathbb{R}^d;\mathbb{C}^n)},
\end{equation*}
hold true, where the constant $C$ is independent of $\e$ and
$\mathbf{u}$.
\end{lemma}

\begin{proof}
The proof follows the ideas of that of Lemma~\ref{lm3.7}; one
just need to make some minor corrections. We introduce the
matrices $P^{(i)}$ as the solutions to (\ref{3.24}), (\ref{3.8})
with $\widehat{A}_1$ replaced by $\widehat{A}_0$. Then
$P^{(i)}\in
\Hinf^1(\mathbb{R}^d;C_{per}^{2+\b}(\overline{\square}))$. By
the first equation in (\ref{1.10}), the analogues of the
relations (\ref{3.23}), (\ref{3.4a}), (\ref{3.4c}) are as
follows
\begin{align*}
&B(\p_\xi)^*\widehat{A}_0=\sum\limits_{i=1}^{d}b_i^*\frac{\p
a_i^*}{\p\xi_i},&& Q:=\sum\limits_{i=1}^{d}\frac{\p
P^{(i)}}{\p\xi_i},
\\
-&B_i^* \widehat{A}_0=\sum\limits_{j=1}^{d}\frac{\p
M_{ij}}{\p\xi_j}+\frac{\p Q}{\p\xi_i},&&\D_\xi
Q=\sum\limits_{i=1}^{d}b_i^* \frac{\p a_i^*}{\p\xi_i}.
\end{align*}
Employing these identities, by analogy with (\ref{3.4d}) we
obtain
\begin{align*}
B(\p_x)^*\widehat{A}_0\mathbf{v}&=
\e\sum\limits_{i,j=1}^{d}\p_j\frac{\p M_{ij}\mathbf{v}}{\p x_i}
+\sum\limits_{i=1}^{d}\frac{\p^2 Q\mathbf{v}}{\p x_i\p\xi_i}
\\
&=\e\sum\limits_{i,j=1}^{d}\p_j\frac{\p M_{ij}\mathbf{v}}{\p
x_i}+\e \sum\limits_{i=1}^{d}\p_i\frac{\p Q\mathbf{v}}{\p x_i}-
\e \sum\limits_{i=1}^{d}\frac{\p^2 Q\mathbf{v}}{\p x_i^2}.
\end{align*}
Now we let
\begin{align*}
&\mathbf{v}_i^{(\e)}(x):=\sum\limits_{j=1}^{d}\frac{\p}{\p x_j}
M_{ji}\left(x,\frac{x}{\e}\right)\mathbf{v}(x)+\frac{\p}{\p x_i}
Q\left(x,\frac{x}{\e}\right)\mathbf{v}(x),
\\
& \mathbf{v}_0^{(\e)}(x):=-\sum\limits_{i=1}^{d}\frac{\p^2}{\p
x_i^2}Q\left(x,\frac{x}{\e}\right)\mathbf{v}(x).
\end{align*}
The aforementioned properties of $P^{(i)}$ yield that the
vector-functions  $\mathbf{v}_i^{(\e)}$ belong to
$L_2(\mathbb{R}^d;\mathbb{C}^n)$ and satisfy the claimed
estimate. To complete the proof, it remains to establish the
same fact for $\mathbf{v}_0^{(\e)}$. In order to do it, it is
sufficient to check that $Q\in
\Hinf^2(\mathbb{R}^d;C^{\b}_{per}(\overline{\square}))$.

Let $Q^{(i)}$ be the solutions of the equations
\begin{equation*}
\D_\xi Q^{(i)}=b_i^*\left(a_i^*-\frac{1}{|\square|}
\int\limits_\square a_i^*(\cdot,\xi)\di\xi\right),
\quad(x,\xi)\in\mathbb{R}^{2d},
\end{equation*}
satisfying the condition (\ref{3.8}). Applying Lemma~\ref{lm3.0}
and differentiating these equations w.r.t. $x_j$, it is easy to
check that $Q^{(i)}\in
\Hinf^2(\mathbb{R}^d;C^{2+\b}_{per}(\overline{\square}))$.
Clearly,
\begin{equation*}
Q=\sum\limits_{i=1}^{d}\frac{\p Q^{(i)}}{\p\xi_i},
\end{equation*}
that implies the desired belonging for $Q$.
\end{proof}

Let $\mathfrak{h}_\e$ be the lower bound of $\mathcal{H}_\e$,
$\mu_\e:=\min\left\{\frac{\mathfrak{h}_\e}{\mathfrak{g}_1},
\frac{\mathfrak{h}_\e}{\mathfrak{g}_2}\right\}$.

\begin{lemma}\label{lm3.8}
Suppose that $\l\in\mathbb{C}\setminus[\mu,+\infty)$, where
$\mu_\e-\mu\geqslant c>0$, and the constant  $c$ is independent
of $\e$. Then the generalized solution
$\mathbf{u}\in\H^1(\mathbb{R}^d;\mathbb{C}^n)$ of
\begin{equation*}
\big(B(\p)^*A_\e B(\p)+a_\e(x,\p)+V_\e-\l G_\e\big)
\mathbf{u}=\mathbf{f}_0+\sum\limits_{i=1}^{d}
\p_{i}\mathbf{f}_i,\quad
\mathbf{f}_i\in L_2(\mathbb{R}^d;\mathbb{C}^n),
\end{equation*}
satisfies the estimate
\begin{equation*}
\|\mathbf{u}\|_{\H^1(\mathbb{R}^d;\mathbb{C}^n)}\leqslant C(\l)
\sum\limits_{i=0}^{d}
\|\mathbf{f}_i\|_{L_2(\mathbb{R}^d;\mathbb{C}^n)},
\end{equation*}
where the constant $C(\l)$ is independent of $\e$ and
$\mathbf{f}_i$.
\end{lemma}

\begin{proof}
Basing on Lemma~\ref{lm2.1}, the
identity
\begin{equation}\label{3.29}
h_\e[\mathbf{u}]-\l
(G_\e\mathbf{u},\mathbf{u})_{L_2(\mathbb{R}^d;\mathbb{C}^n)}=
(\mathbf{f}_0,\mathbf{u})_{L_2(\mathbb{R}^d;\mathbb{C}^n)} -
\sum\limits_{i=1}^{d}
(\mathbf{f}_i,\p_{i}\mathbf{u})_{L_2(\mathbb{R}^d;\mathbb{C}^n)} 
\end{equation}
and  (\ref{1.6a}) one can prove that
\begin{equation}\label{3.28}
\|\mathbf{u}\|_{\H^1(\mathbb{R}^d;\mathbb{C}^n)}\leqslant C(\l)
\left( \sum\limits_{i=0}^{d}
\|\mathbf{f}_i\|_{L_2(\mathbb{R}^d;\mathbb{C}^n)}+
\|\mathbf{u}\|_{L_2(\mathbb{R}^d;\mathbb{C}^n)}\right),
\end{equation}
where the constant $C$ is independent of $\e$ and
$\mathbf{f}_i$. The first term in the left hand side of
(\ref{3.29}) being real, this identity implies that
\begin{equation*}
-\IM\l(G_\e\mathbf{u},\mathbf{u})_{L_2(\mathbb{R}^d;\mathbb{C}^n)}=
\IM(\mathbf{f}_0,\mathbf{u})_{L_2(\mathbb{R}^d;\mathbb{C}^n)}
-\IM\sum\limits_{i=1}^{d}
(\mathbf{f}_i,\p_{i}\mathbf{u})_{L_2(\mathbb{R}^d;\mathbb{C}^n)}. 
\end{equation*}
If $\IM\l\not=0$, this identity and (\ref{1.6a}) yields
\begin{equation}\label{3.6a}
\|\mathbf{u}\|_{L_2(\mathbb{R}^d;\mathbb{C}^n)}^2\leqslant
\d\|\mathbf{u}\|_{\H^1(\mathbb{R}^d;\mathbb{C}^n)}^2+C(\d,\l)
\sum\limits_{i=0}^{d}
\|\mathbf{f}_i\|_{L_2(\mathbb{R}^d;\mathbb{C}^n)}^2,
\end{equation}
where the number $\d$ can be chosen anyhow small, and the
constant $C$ is independent of $\e$ and $\mathbf{f}_i$. If
$\l\in(-\infty,\mu)$, the last estimate is valid as well that
follows from (\ref{3.29}) and the inequality
\begin{align*}
h_\e[\mathbf{u}]-&\l(G_\e
\mathbf{u},\mathbf{u})_{L_2(\mathbb{R}^d;\mathbb{C}^n)}
\geqslant \mathfrak{h}_\e
\|\mathbf{u}\|_{L_2(\mathbb{R}^d;\mathbb{C}^n)}^2-\mu (G_\e
\mathbf{u},\mathbf{u})_{L_2(\mathbb{R}^d;\mathbb{C}^n)}
\\
&\geqslant(\mathfrak{h}_\e-\mu \mathfrak{g})
\|\mathbf{u}\|_{L_2(\mathbb{R}^d;\mathbb{C}^n)}^2\geqslant
(\mu_\e-\mu)\mathfrak{g}
\|\mathbf{u}\|_{L_2(\mathbb{R}^d;\mathbb{C}^n)}^2\geqslant
c\mathfrak{g} \|\mathbf{u}\|_{L_2(\mathbb{R}^d;\mathbb{C}^n)}^2,
\end{align*}
where
\begin{align*}
\mathfrak{g}=\left\{
\begin{aligned}
&\mathfrak{g}_2, && \text{если} &&\mu\geqslant 0,
\\
&\mathfrak{g}_1, && \text{если} &&\mu<0.
\end{aligned} \right.
\end{align*}
Now the estimates (\ref{3.28}), (\ref{3.6a}) lead us to the
statement of lemma.
\end{proof}

\begin{proof}[Proof of Theorem~\ref{th1.2} for non-real
$\l$] Let $\mathbf{f}\in L_2(\mathbb{R}^d;\mathbb{C}^n)$,
\begin{align*}
&\mathbf{u}^{(\e)}:=(\mathcal{H}_\e-\l
G_\e)^{-1}\mathbf{f},\quad \mathbf{u}^{(0)}:=(\mathcal{H}_0-\l
G_0)^{-1}\mathbf{f},
\\
&\mathbf{u}^{(1)}(x,\xi):=
\left(\L_1(x,\xi)B(\p)+\L_0(x,\xi)\right)\mathbf{u}^{(0)}(x),
\quad \widehat{\mathbf{u}}^{(\e)}(x):=\mathbf{u}^{(0)}(x)+\e
\mathbf{u}^{(1)}\left(x,\frac{x}{\e}\right).
\end{align*}
It is obvious that
\begin{equation}\label{3.27}
\begin{aligned}
\big(B&(\p)^*A_\e B(\p)+a_\e(x,\p)+V_\e-\l G_\e\big)\big(
\mathbf{u}^{(\e)}-\widehat{\mathbf{u}}^{(\e)}\big)
\\
&=\mathbf{f}- \big(B(\p)^*A_\e B(\p)+a_\e(x,\p)+V_\e-\l G_\e
\big)\widehat{\mathbf{u}}^{(\e)}
\\
&=(\mathcal{H}_0-\l G_0)\mathbf{u}^{(0)}- \big(B(\p)^*A_\e
B(\p)+a_\e(x,\p)+V_\e-\l G_\e\big)
\widehat{\mathbf{u}}^{(\e)}=:\mathbf{F}^{(\e)}.
\end{aligned}
\end{equation}
Let us evaluate the function $\mathbf{F}^{(\e)}$. Taking into
account the identities (\ref{3.0a}) and the second equation in
(\ref{1.10}) we obtain
\begin{equation}\label{3.6b}
\begin{aligned}
&B(\p)^*A B(\p)\widehat{\mathbf{u}}^{(\e)}= B(\p_x)^*A
B(\p)\mathbf{u}^{(0)}+ B(\p_x)^*A B(\p_\xi)\mathbf{u}^{(1)}
\\
&+\e B(\p)^*A B(\p_x)\mathbf{u}^{(1)}+\e^{-1} B(\p_\xi)^*A
B(\p)\mathbf{u}^{(0)} +\e^{-1}B(\p_\xi)^*A
B(\p_\xi)\mathbf{u}^{(1)}
\\
&=B(\p_x)^*(A+B(\p_\xi)\L_1)B(\p)\mathbf{u}^{(0)}+ B(\p_x)^*A
B(\p_\xi)\L_0\mathbf{u}^{(0)}
\\
&+\e B(\p)^*A B(\p_x)\mathbf{u}^{(1)}+\e^{-1} B(\p_\xi)^*A
B(\p_\xi)\L_0\mathbf{u}^{(0)},
\\
&a_\e(x,\p) \widehat{\mathbf{u}}^{(\e)}= \sum\limits_{i=1}^{d}
\left(a_i\frac{\p}{\p x_i}b_i \mathbf{u}^{(0)}-b_i^*
\frac{\p}{\p x_i}a_i^* \mathbf{u}^{(0)}\right)- \e^{-1}
\sum\limits_{i=1}^{d}b_i^*\frac{\p
a_i^*}{\p\xi_i}\mathbf{u}^{(0)}
\\
&+\e \sum\limits_{i=1}^{d} \left(a_i\frac{\p}{\p x_i}b_i
\mathbf{u}^{(1)}+ \frac{\p b_i^* }{\p x_i} a_i^*
\mathbf{u}^{(1)}-\p_i b_i^* a_i^*
\mathbf{u}^{(1)}\right)+\sum\limits_{i=1}^{d} a_i b_i \frac{\p
\mathbf{u}^{(1)}}{\p\xi_i}.
\end{aligned}
\end{equation}
Here the arguments of all functions except
$\mathbf{u}^{(0)}(x)$, $\mathbf{u}^{(\e)}(x)$ and
$\widehat{\mathbf{u}}^{(\e)}(x)$ are
$\left(x,\frac{x}{\e}\right)$. Integrating by parts and
employing the equations (\ref{1.10}) one can make sure that
\begin{equation}
\begin{aligned}
&\int\limits_\square (B(\p_\xi)\L_0)^*A \di\xi=
\int\limits_\square \L_0^* B(\p_\xi)^* A\di\xi=
-\int\limits_\square \L_0^* B(\p_\xi)^* A B(\p_\xi)\L_1\di\xi
\\
&=-\int\limits_\square (B(\p_\xi)^* A
B(\p_\xi)\L_0)^*\L_1\di\xi=
-\sum\limits_{i=1}^{d}\int\limits_\square \frac{\p a_i}{\p\xi_i}
b_i \L_1\di\xi=\sum\limits_{i=1}^{d}\int\limits_\square a_i b_i
\frac{\p\L_1}{\p\xi_i}\di\xi,
\\
&\int\limits_\square  \big(B(\p_\xi)\L_0\big)^* A
B(\p_\xi)\L_0\di\xi=\int\limits_\square \big(A
B(\p_\xi)\L_0\big)^* B(\p_\xi)\L_0\di\xi
\\
&= \int\limits_\square \big(B(\p_\xi)^*A
B(\p_\xi)\L_0\big)^*\L_0\di\xi
=\sum\limits_{i=1}^{d}\int\limits_\square \frac{\p a_i}{\p\xi_i}
b_i \L_0\di\xi=-\sum\limits_{i=1}^{d}\int\limits_\square a_i b_i
\frac{\p\L_0}{\p\xi_i}\di\xi.
\end{aligned}\label{3.9a}
\end{equation}
Here the arguments of all matrices are $(x,\xi)$. These
identities and (\ref{1.14}) yield
\begin{align*}
&A_1(x,\p):=\frac{1}{|\square|}B(\p)^*\int\limits_\square
A(x,\xi) B(\p_\xi)\L_0(x,\xi)\di\xi
\\
&\hphantom{A_1(x,\p):=}+\left(\frac{1}{|\square|}\int\limits_\square
a_i(x,\xi) b_i(x)\frac{\p\L_1}{\p\xi_i} (x,\xi)\di\xi\right)
B(\p)+\frac{1}{|\square|}\int\limits_\square a(x,\xi,\p)\di\xi,
\\
&A_0(x):=\frac{1}{|\square|}\sum\limits_{i=1}^{d}
\int\limits_\square a_i(x,\xi)
b_i(x)\frac{\p\L_0}{\p\xi_i}(x,\xi)\di\xi
+\frac{1}{|\square|}\int\limits_\square V(x,\xi)\di\xi.
\end{align*}
Bearing in mind  these relations, (\ref{3.6b}), (\ref{1.10}),
and the definition of $\mathbf{u}^{(1)}$, we obtain
\begin{align*}
&\mathbf{F}^{(\e)}=\mathbf{F}_1^{(\e)}+\mathbf{F}_2^{(\e)}+
\mathbf{F}_3^{(\e)},\quad \mathbf{F}_1^{(\e)}=-B(\p_x)^*
\widehat{A}_1 B(\p)\mathbf{u}^{(0)}-B(\p_x)^*\widehat{A}_0
\mathbf{u}^{(0)},
\\
&\mathbf{F}_2^{(\e)}= \sum\limits_{i=1}^{d}
\left(\frac{1}{|\square|} \int\limits_\square
a_i(\cdot,\xi)\di\xi-a_i\right)\frac{\p}{\p x_i} b_i
\mathbf{u}^{(0)}
\\
&\hphantom{\mathbf{F}_2^{(\e)}=}-\sum\limits_{i=1}^{d}
b_i^*\frac{\p}{\p x_i}\left(\frac{1}{|\square|}
\int\limits_\square a_i^*(\cdot,\xi)\di\xi-a_i^*\right)
\mathbf{u}^{(0)}
\\
&\hphantom{\mathbf{F}_2^{(\e)}=}+\sum\limits_{i=1}^{d}
\left(\frac{1}{|\square|} \int\limits_{\square} a_i(\cdot,\xi)
b_i(\cdot)\frac{\p\L_1}{\p\xi_i}(\cdot,\xi)\di\xi -a_i
b_i\frac{\p\L_1}{\p\xi_i}\right)B(\p) \mathbf{u}^{(0)}\nonumber
\\
&\hphantom{\mathbf{F}_2^{(\e)}=}+\sum\limits_{i=1}^{d}
\left(\frac{1}{|\square|} \int\limits_\square a_i(\cdot,\xi)
b_i(\cdot)\frac{\p\L_0}{\p\xi_i}(\cdot,\xi)\di\xi-a_i
b_i\frac{\p\L_0}{\p\xi_i}\right)\mathbf{u}^{(0)}
\\
&\hphantom{\mathbf{F}_2^{(\e)}=}
+\left(\frac{1}{|\square|}\int\limits_\square
V(\cdot,\xi)\di\xi-V\right)\mathbf{u}^{(0)}+\l(G-G_0)
\mathbf{u}^{(0)},
\\
&\mathbf{F}_3^{(\e)}=-\e B(\p)^* A B(\p_x)\mathbf{u}^{(1)}-\e
\sum\limits_{i=1}^{d} \left(a_i\frac{\p}{\p x_i }b_i+\frac{\p
b_i^*}{\p x_i} a_i^* -\p_i
b_i^*a_i^*\right)\mathbf{u}^{(1)}-\e(V-\l G) \mathbf{u}^{(1)},
\end{align*}
where the arguments of the functions are
$\big(x,\frac{x}{\e}\big)$. The belonging $\L_i\in
\Hinf^1(\mathbb{R}^d;C_{per}^{2+\b}(\overline{\square}))$ and
the inequality
\begin{equation}\label{3.19}
\|\mathbf{u}^{(0)}\|_{\H^2(\mathbb{R}^d;\mathbb{C}^n)}\leqslant
C\|\mathbf{f}\|_{L_2(\mathbb{R}^d;\mathbb{C}^n)},
\end{equation}
imply that
\begin{equation*}
\left\|B(\p_x)\mathbf{u}^{(1)}
\left(x,\frac{x}{\e}\right)\right\|_{L_2(\mathbb{R}^d;\mathbb{C}^n)}
\leqslant C\|\mathbf{f}\|_{L_2(\mathbb{R}^d;\mathbb{C}^n)}.
\end{equation*}
Hereinafter till the end of the proof by $C$ we denote
inessential constants independent of $\e$ and $\mathbf{f}$.
Taking into account the last estimate we conclude that the
function $\mathbf{F}_3^{(\e)}$ can be represented as
\begin{equation}\label{3.26a}
\mathbf{F}_3^{(\e)}=\e \sum\limits_{i=1}^{d}
\frac{\p\mathbf{f}_{3,i}^{(\e)}}{\p x_i}+\e
\mathbf{f}_{3,0}^{(\e)},\quad
\|\mathbf{f}_{3,i}^{(\e)}\|_{L_2(\mathbb{R}^d;\mathbb{C}^n)}
\leqslant C\|\mathbf{f}\|_{L_2(\mathbb{R}^d;\mathbb{C}^n)},\quad
i=0,\ldots,d.
\end{equation}
The formula for $\mathbf{F}_2^{(\e)}$ implies immediately that
this function is a sum of terms each of them satisfies the
hypothesis of Lemma~\ref{lm3.6}. Hence by this lemma we have
\begin{equation*}
\mathbf{F}_2^{(\e)}=\e \sum\limits_{i=1}^{d}
\frac{\p\mathbf{f}_{2,i}^{(\e)}}{\p x_i}+\e
\mathbf{f}_{2,0}^{(\e)},\quad
\|\mathbf{f}_{2,i}^{(\e)}\|_{L_2(\mathbb{R}^d;\mathbb{C}^n)}
\leqslant C\|\mathbf{f}\|_{L_2(\mathbb{R}^d;\mathbb{C}^n)},\quad
i=0,\ldots,d.
\end{equation*}
where the constant $C$ is independent of $\e$ and $\mathbf{f}$.
These identities, Lemma~\ref{lm3.7} and (\ref{3.19}),
(\ref{3.26a}) yield
\begin{equation*}
\mathbf{F}^{(\e)}=\e \sum\limits_{i=1}^{d}
\frac{\p\mathbf{f}_{i}^{(\e)}}{\p x_i}+\e
\mathbf{f}_{0}^{(\e)},\quad
\|\mathbf{f}_{i}^{(\e)}\|_{L_2(\mathbb{R}^d;\mathbb{C}^n)}
\leqslant C\|\mathbf{f}\|_{L_2(\mathbb{R}^d;\mathbb{C}^n)},\quad
i=0,\ldots,d.
\end{equation*}
We substitute this representation into (\ref{3.27}) and by
Lemma~\ref{lm3.8} we arrive at the estimate
\begin{equation*}
\|\mathbf{u}^{(\e)}-
\widehat{\mathbf{u}}^{(\e)}\|_{\H^1(\mathbb{R}^d;\mathbb{C}^n)}
\leqslant C\sum\limits_{i=0}^{d}
\|\mathbf{f}_i^{(\e)}\|_{L_2(\mathbb{R}^d;\mathbb{C}^n)}
\leqslant C\e\|\mathbf{f}\|_{L_2(\mathbb{R}^d;\mathbb{C}^n)}.
\end{equation*}
This leads us immediately to the latter estimate in
(\ref{1.15}).  Employing this estimate and Lemma~\ref{lm3.5} we
obtain
\begin{align*}
\|\mathbf{u}^{(\e)}-&
\mathbf{u}^{(0)}\|_{L_2(\mathbb{R}^d;\mathbb{C}^n)}\leqslant
\\
&\leqslant C \|\mathbf{u}^{(\e)}-
\widehat{\mathbf{u}}^{(\e)}\|_{L_2(\mathbb{R}^d;\mathbb{C}^n)}+C
\e\|\mathcal{L}_\e(\mathcal{H}_0-\l G_0)^{-1} \mathbf{f}
\|_{L_2(\mathbb{R}^d;\mathbb{C}^n)} \leqslant
C\e\|\mathbf{f}\|_{L_2(\mathbb{R}^d;\mathbb{C}^n)}.
\end{align*}
The former estimate in (\ref{1.15}) is proven.
\end{proof}

Corollary~\ref{th1.1} is implied by the former estimate in
(\ref{1.15}) with $G=G_0=E_n$ and
Theorems~V\!I\!I\!I.23,~V\!I\!I\!I.24 in \cite[Ch. V\!I\!I\!I,
\S 7]{RS}.

\begin{proof}[Proof of Theorem~\ref{th1.2} for
$\l\in(-\infty,\mu_0)$] By Corollary~\ref{th1.1} the lower bound
of $\mathcal{H}_\e$ converges to that of $\mathcal{H}_0$. Hence,
$\mu_\e\to\mu_0$ as $\e\to+0$, and thus, for sufficiently small
$\e$ the number $\l\in(-\infty,\mu_0)$ satisfies
Lemma~\ref{lm3.8}. It follows from (\ref{1.6a}) that this
estimate holds true for $G_0$ as well. It is also easy to check
that the estimate (\ref{3.19}) is valid. In view of these facts
it is clear that all the arguments in the proof of the theorem
for  $\IM\l\not=0$ remain valid in the case
$\l\in(-\infty,\mu_0)$, if $\e$ is small enough. It proves the
estimates (\ref{1.15}) in the latter case as well.
\end{proof}

\section{Examples}

In the section we give examples of certain operators to which
the results of the previous sections can be applied.

Our first example is
\begin{equation}\label{6.1}
\mathcal{H}_\e:=\sum\limits_{i,j=1}^{d} \left(-\p_i
+\mathsf{a}_{i,\e}^*\right)\mathsf{g}^{ij}_\e \left(\p_j
+\mathsf{a}_{j,\e}\right)+\mathsf{v}_\e,
\end{equation}
where
\begin{align*}
&\mathsf{g}^{ij}=\mathsf{g}^{ij}(x,\xi)\in
\Hinf^1(\mathbb{R}^d;C^{1+\b}_{per}(\overline{\square}))\cap
C_*^2(\mathbb{R}^d;C^{\b}_{per}(\overline{\square})),
\\
&\mathsf{a}_i=\mathsf{a}_i(x,\xi)\in
\Hinf^1(\mathbb{R}^d;C^{1+\b}_{per}(\overline{\square}))\cap
C_*^2(\mathbb{R}^d;C^{\b}_{per}(\overline{\square})),
\\
&\mathsf{v}=\mathsf{v}(x,\xi)\in
\Hinf^1(\mathbb{R}^d;C^{\b}_{per}(\overline{\square}))
\end{align*}
are $\square$-periodic matrices of the size $n\times n$.
Moreover, the identities $\mathsf{v}=\mathsf{v}^*$,
$(\mathsf{g}^{ij})^*=\mathsf{g}^{ji}$ are supposed to be valid
as well as
\begin{equation*}
c_1\sum\limits_{i=1}^{d} \|\mathbf{w}_i\|_{\mathbb{C}^n}^2
\leqslant \sum\limits_{i,j=1}^{d}
(\mathsf{g}^{ij}\mathbf{w}_j,\mathbf{w}_i)\leqslant
c_2\sum\limits_{i=1}^{d} \|\mathbf{w}_i\|_{\mathbb{C}^n}^2
\end{equation*}
for all $\mathbf{w}_i\in \mathbb{C}^n$,
$(x,\xi)\in\mathbb{R}^{2d}$, where $c_1$, $c_2$ are constants.
The operator (\ref{6.1}) can be written as (\ref{1.5}); let us
indicate the corresponding choice of  $A$, $a_i$, $b_i$, $V$.

We let $m=nd$ and choose the matrices $A$ and $B(\z)$ as
\begin{equation*}
B(\z):=
\begin{pmatrix}
\z_1 E_n
\\
\z_2 E_n
\\
\vdots
\\
\z_d E_n
\end{pmatrix},
\quad A:=
\begin{pmatrix}
\mathsf{g}^{11} & \mathsf{g}^{12} & \ldots & \mathsf{g}^{1d}
\\
\mathsf{g}^{21} & \mathsf{g}^{22} & \ldots & \mathsf{g}^{2d}
\\
\vdots & \vdots &        & \vdots
\\
\mathsf{g}^{d1} & \mathsf{g}^{d2} & \ldots & \mathsf{g}^{dd}
\end{pmatrix}.
\end{equation*}
The matrices $a_i$, $b_i$ and $V$ are introduced as
\begin{equation*}
a_i:=\sum\limits_{j=1}^{d} \mathsf{a}_j^* \mathsf{g}^{ji}, \quad
b_i:=E_n,\quad V:=\mathsf{v}+\sum\limits_{i,j=1}^{d}
\mathsf{a}_i^* \mathsf{g}^{ij} \mathsf{a}_j.
\end{equation*}
It is easy to check that in this case the operator in
(\ref{1.5}) coincides with the operator in (\ref{6.1}).  Many
operators of the mathematical physics are the particular cases
of (\ref{6.1}); let us mention some of them.

If we let $\mathsf{a}_i:=0$, $\mathsf{g}^{ij}:=E_n$, the
operator (\ref{6.1}) is a matrix Schr\"odinger operator. The
case $\mathsf{g}^{ij}\not\equiv E_n$ can be considered as the
matrix Schr\"odinger operator with a metric. If, in addition,
$\mathsf{v}=0$, we arrive at the operator of the elasticity
theory; one just needs to assume additional symmetricity
conditions for the coefficients of the matrix $\mathsf{g}^{ij}$
(see, for instance, \cite[Ch. 3]{PChSh}).

In the case $n=1$, $\mathsf{a}_i:=\iu \mathsf{A}_i$,
$\mathsf{A}_i$ are real-valued function, the operator
(\ref{6.1}) describes the magnetic Schr\"odinger operator. The
components of the magnetic potential are the functions
$\mathsf{A}_i$; the function $\mathsf{v}$ is the electric
potential. As above, the functions $\mathsf{g}^{ij}$ correspond
to the metric.

One more example is the two- and three-dimensional Pauli
operator. We deal with this operator, if $d=2$ or $d=3$, $n=2$,
$\mathsf{a}_i:=\iu\mathsf{A}_i E_n$, $\mathsf{A}_i$ are
real-valued function,
\begin{align*}
&\mathsf{v}:=\si_3 B, \quad B=\frac{\p \mathsf{A}_2}{\p
x_1}-\frac{\p \mathsf{A}_1}{\p x_2}, \quad \text{if} \quad d=2,
\\
&\mathsf{v}:=\si_1 B_1+\si_2 B_2+\si_3 B_3, \quad
(B_1,B_2,B_3)=\rot (\mathsf{A}_1,\mathsf{A}_2,\mathsf{A}_3),
\quad
\text{if}\quad d=3, 
\\
& \si_1:=
\begin{pmatrix}
0 & 1
\\
1 & 0
\end{pmatrix},
\quad \si_2:=
\begin{pmatrix}
0 & -\iu
\\
\iu & 0
\end{pmatrix},
\quad \si_3:=
\begin{pmatrix}
1 & 0
\\
0 & -1
\end{pmatrix}.
\end{align*}
The case $\mathsf{g}^{ij}=E_n$ corresponds to the usual Pauli
operator; if $\mathsf{g}^{ij}\not\equiv E_n$, we obtain the
Pauli operator with metric. One can add an additional term to
the potential $\mathsf{v}$ given above. In this case we have
Pauli operator with potential.

In the examples given all the results of Theorem~\ref{th1.2} and
Corollary~\ref{th1.1} are applicable. The homogenized operator
is given by the general formulas (\ref{1.13}), (\ref{1.14}).
This is why we will not repeat these formulas for the particular
cases described.

The presence of the matrix $G_\e$ in the estimates (\ref{1.15})
allow us to wide the class of the examples. In order to do it,
we employ the ideas of papers \cite{BS}, \cite{BS2}, \cite{BS5}.

Let $\mathsf{f}=\mathsf{f}(x,\xi)\in
\Hinf^1(\mathbb{R}^d;C^{2+\b}_{per}(\overline{\square}))$ be a
positive matrix of the size $n\times n$ such that the inverse
matrix is uniformly bounded. We consider the operator
$\widetilde{\mathcal{H}}_\e:=\mathsf{f}_\e^* \mathcal{H}_\e
\mathsf{f}_\e$, where $\mathcal{H}_\e$ is from (\ref{1.5}). It
is clear that
\begin{equation*}
\mathsf{f}_\e(\widetilde{\mathcal{H}}_\e-\l
G_\e)^{-1}\mathsf{f}_\e^*=(\mathcal{H}_\e-\l
\widetilde{G}_\e)^{-1},\quad \widetilde{G}=(\mathsf{f}^*)^{-1}G
\mathsf{f}^{-1}.
\end{equation*}
It allows us to approximate the generalized resolvent of the
operator $\widetilde{\mathcal{H}}_\e$,
\begin{align*}
&\|(\widetilde{\mathcal{H}}_\e-\l G_\e)^{-1}-\mathsf{f}_\e^{-1}
(\mathcal{H}_0-\l G_0 )^{-1}(\mathsf{f}_\e^*)^{-1}\|_{L_2\to
L_2}\leqslant C\e,
\\
&\|\mathsf{f}_\e(\widetilde{\mathcal{H}}_\e-\l G_\e)^{-1}-
(\I+\e \mathcal{L}_\e)(\mathcal{H}_0-\l
G_0)^{-1}(\mathsf{f}_\e^*)^{-1}\|_{L_2\to \H^1}\leqslant C\e.
\end{align*}
Let us introduce now the operator $\mathcal{H}_\e$ by
(\ref{6.1}); the corresponding operator
$\widetilde{\mathcal{H}}_\e$ is determined by the same formula
but with the coefficients replaced by
\begin{align*}
\widetilde{\mathsf{g}}^{ij}:=\mathsf{f}^*\mathsf{g}^{ij}\mathsf{f},
\quad \widetilde{\mathsf{a}}_i:=\mathsf{f}^{-1}\left(\frac{\p
\mathsf{f}}{\p x_i}+ \e^{-1}\frac{\p
\mathsf{f}}{\p\xi_i}\right)+\mathsf{f}^{-1} \mathsf{a}_i
\mathsf{f},\quad \widetilde{\mathsf{v}}:=\mathsf{f}^* \mathsf{v}
\mathsf{f}.
\end{align*}
As it follows from these formulas, the coefficients of the
operator $\widetilde{\mathcal{H}}_\e$ can increase as $\e\to+0$.
It allows us to extend the results of the paper to the certain
class of the operators with fast oscillating coefficients
increasing as  $\e\to+0$.

In conclusion we also observe that in the case $\mathsf{a}_i=0$,
$v=0$ the operator $\widetilde{\mathcal{H}}_\e$ was the main
object of the study in \cite{BS}--\cite{BS5}; as it has been
already mentioned in the beginning of the paper, under the
essentially weaker assumptions for the coefficients. In the
cited the authors gave a great number of interesting examples
for such operators. Our results can extended to these examples
as well. The novelty will be the dependence of the coefficients
of slow variable and the asymptotics expansions for the
eigenelements.

\section*{Acknowledgements}

I am very grateful to T.A.~Suslina for the attention to the
paper, the discussion on the results, and many valuable remarks.

The work is supported in parts by RFBR (07-01-00037) and by the
Czech Academy of Sciences and Ministry of Education, Youth and
Sports (LC06002). The author is also supported by \emph{Marie
Curie International Fellowship} within 6th European Community
Framework Programm (MIF1-CT-2005-006254). The author gratefully
acknowledges the support from Deligne 2004 Balzan prize in
mathematics.

\end{document}